\documentclass[sigconf, screen]{acmart}
\setcopyright{none}
\renewcommand\footnotetextcopyrightpermission[1]{}
\AtBeginDocument{%
  }

\begin{document}

\title{Ink Spiral: Symbolic Transformation from The Thinker to the Four Gentlemen}

\author{Lingyu Peng}
\email{lingyupeng6@163.com}
\orcid{0009-0000-5964-3340}
\affiliation{%
  \institution{Future Design School, Harbin Institute of Technology, Shenzhen}
  \city{Shenzhen}
  \state{Guangdong}
  \country{China}
}

\author{Wenbo Lu}
\email{24s039099@stu.hit.edu.cn}
\orcid{0009-0002-6218-9745}
\affiliation{%
  \institution{Future Design School, Harbin Institute of Technology, Shenzhen}
  \city{Shenzhen}
  \state{Guangdong}
  \country{China}
}

\author{Liying Long}
\email{longliying1@163.com}
\orcid{0009-0006-1228-847X}
\affiliation{%
  \institution{Future Design School, Harbin Institute of Technology, Shenzhen}
  \city{Shenzhen}
  \state{Guangdong}
  \country{China}
}

\author{Qingchuan Li}
\authornote{Qingchuan Li is the corresponding author.}
\email{liqingchuan@hit.edu.cn}
\orcid{0000-0001-9915-2589}
\affiliation{%
  \institution{Future Design School, Harbin Institute of Technology, Shenzhen}  
  \city{Shenzhen}
  \state{Guangdong}
  \country{China}
}

\renewcommand{\shortauthors}{Peng et al.}

\begin{abstract}
Western art has regarded \textit{The Thinker} as a symbol of rational contemplation, while Eastern aesthetics has taken the Four Gentlemen—plum, orchid, bamboo, and chrysanthemum—as symbols of moral and spiritual cultivation. This paper presents Ink Spiral, a video installation that links these traditions through AI-generated ink imagery. By transforming a rotating sculpture of \textit{The Thinker} into the Four Gentlemen across thousands of frames, the work shifts between three-dimensional sculpture and two-dimensional ink, human introspection and natural symbolism. Ink Spiral turns fixed cultural icons into a fluid dialogue, inviting audiences to perceive cross-cultural connection as a living, ambiguous, and endlessly interpretable creative state.
\end{abstract}


\begin{teaserfigure}
  \centering
  \includegraphics[width=0.9\textwidth]{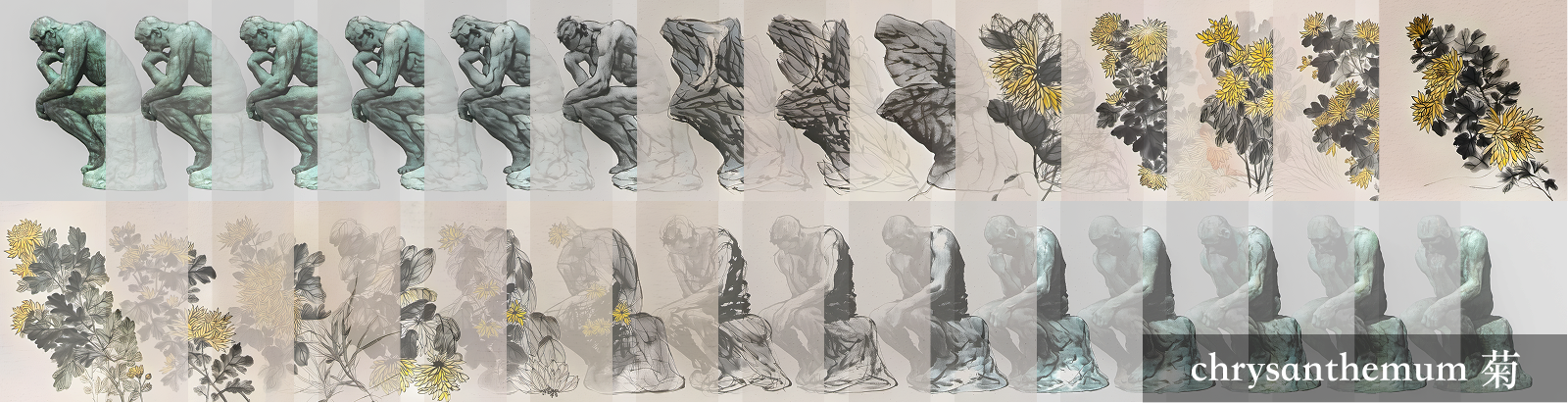}
  \caption{Key frames showing the gradual transformation of \textit{The Thinker} through rotation into a chrysanthemum ink painting and back into sculptural form.}
  \Description{teaser}
  \label{fig:teaser}
\end{teaserfigure}

\received{20 February 2007}
\received[revised]{12 March 2009}
\received[accepted]{5 June 2009}

\maketitle


\section{Introduction}
Ink Spiral is a video installation that explores the encounter between Western rational thought and Eastern philosophical aesthetics through AI-generated ink imagery. Centered on a rotating sculpture created by Rodin’s \textit{The Thinker}, the work gradually transforms this iconic figure into the Four Gentlemen—plum, orchid, bamboo, and chrysanthemum—as it turns. Through this continuous morphing process, the installation connects the introspective qualities associated with \textit{The Thinker} with Eastern traditions that express moral character and existential values through symbolic natural forms.

By coordinating transformation across image, style, space, and time, Ink Spiral constructs an ambiguous visual relation between these two symbolic systems. As one dissolves into the other without a fixed boundary, the work resists stable classification between sculpture and ink painting, Western and Eastern symbolism, inviting viewers to reflect on possible connections between them.
\noindent\textbf{Project video:} \url{https://vimeo.com/1185853274?share=copy&fl=sv&fe=ci}
\section{Conceptual Construction}
This section focuses on the two symbolic elements at the center of Ink Spiral, showing how they are reworked into the conceptual basis of the installation.

\subsection{\textit{The Thinker}}
Rodin’s \textit{The Thinker} has long been regarded as a canonical image of introspection, reflective subjectivity, and the search for human existence in Western art \cite{rodin2012rodin}. It represents a distinctly Western mode of thought in which the body carries inward reflection and rational consciousness. Its significance also lies in the language of sculpture, where bodily form, posture, and tension give visible shape to a rational understanding of the human condition \cite{winckelmann1767reflections}.

In Ink Spiral, we take \textit{The Thinker} as the conceptual point of departure for the work. As an intuitive and culturally legible symbol, it offers viewers an accessible point of entry into the installation. At the same time, its sculptural form invites observation from multiple angles, allowing different postures, contours, and tensions to stimulate further reflection and association as the image enters a process of generative transformation.

\subsection{the Four Gentlemen}
The Four Gentlemen, namely plum, orchid, bamboo, and chrysanthemum, occupy an important place in Chinese art and aesthetic thought as symbolic forms through which moral character and reflections on human existence are expressed \cite{sullivan2008arts, bush2012chinese}.They reflect a mode of thought in which people actively turn to natural plants to articulate personal reflection and spiritual meaning. Their significance is closely connected to the language of ink painting. In Eastern ink painting, emphasis is placed on the expression of spirit resonance and the abstraction of form, through which meaning is conveyed by brushwork, atmosphere, and suggestion \cite{harrist1999ink}.

In Ink Spiral, we treat the Four Gentlemen as symbols of another mode of thinking. Through natural forms and their cultural meanings, they offer an alternative way of approaching the self. Their seasonal associations also introduce temporal progression and a sense of living dynamism, allowing transformation to unfold gradually across the video.

\subsection{Constructing Visual and Conceptual Tensions through Video}
Building on the two symbolic systems discussed above, Ink Spiral develops its core concept through a set of visual and philosophical correspondences. These correspondences are not treated as fixed oppositions, but as tensions that unfold gradually through perception and interpretation. Video is therefore essential to the work, as it renders transformation perceptible through duration, continuity, and progressive change rather than static comparison. The key correspondences are summarized in Table~\ref{tab:inkspiral-correspondences}.

\begin{table}[H]
\centering
\footnotesize
\caption{Tensions in Ink Spiral}
\label{tab:inkspiral-correspondences}
\setlength{\tabcolsep}{4pt}
\renewcommand{\arraystretch}{1}
\begin{tabular}{p{0.4\linewidth} c p{0.4\linewidth}}
\toprule
\textbf{Ink / Plant Side} & \textbf{vs.} & \textbf{Sculpture / Human Side} \\
\midrule
Atmospheric and suggestive ink imagery & \textbf{vs.} & Precise and detailed sculptural form \\
Two-dimensional pictorial surface & \textbf{vs.} & Three-dimensional bodily presence \\
Plant motifs as symbolic objects & \textbf{vs.} & Human figure as reflective subject \\
Seasonal change and cyclical time & \textbf{vs.} & Spatial variation through rotation \\
Dynamic vitality of living growth & \textbf{vs.} & Static and solid sculptural presence \\
\bottomrule
\end{tabular}
\end{table}

Video is particularly suited to constructing these tensions, as it can register not only changes in visual content but also the unfolding of spatial and temporal relations.

\section{Implementation}
GenAI is particularly suitable for Ink Spiral because it can generate intermediate states between artistic forms that are difficult to create manually. In this work, the transformation from sculpture to ink painting involves not only a stylistic shift, but also a gradual change in form, symbolism, and perception. Drawing on the idea of the ambiguity of process \cite{sivertsen2024machine}, we treat this indeterminacy as a productive quality that expands the imaginative space between the two philosophical systems.

To realize this process, we first built a dataset of more than 1,000 ink paintings based on the Four Gentlemen, each depicting a single motif and manually annotated by category. We then applied LoRA fine-tuning to a pretrained diffusion model to capture key features of ink painting. To preserve visual coherence during transformation, we used a three-branch ControlNet conditioned on blurred tile-based appearance guidance, depth maps, and segmentation masks \cite{zhang2023adding}. This anchored generation to the rotating sculpture’s silhouette while enabling a controlled transition toward plant morphology and ink expression. Implemented as a frame-by-frame pipeline, the process produced more than 5,000 images, each functioning as an intermediate state within the metamorphosis. As the rotation progressed, the sequence moved from sculptural form to stylistic modulation, then to the emergence of plant morphology, and finally to ink-based growth.

\section{Conclusion}
GenAI is particularly suitable for Ink Spiral because it can produce intermediate states between artistic forms that are otherwise difficult to create manually. In this work, the transformation from sculpture to ink painting involves not only a shift in style, but also a gradual change in form, symbolism, and perception. Drawing on the idea of ambiguity of process, we treat this indeterminacy as a productive quality that expands the imaginative space between the two philosophical systems.

\begin{acks}
This study was funded by the Guangdong Featured Innovation Project in Higher Education (Grant No. 2025WTSCX115).
\end{acks}

\bibliographystyle{ACM-Reference-Format}
\bibliography{main}

@misc{harrist1999ink,
  title={Ink Plum: The Making of a Chinese Scholar-painting Genre},
  author={Harrist Jr, Robert E},
  year={1999},
  publisher={JSTOR}
}

@book{winckelmann1767reflections,
  title={Reflections on the painting and sculpture of the Greeks... Translated... by Henry Fusseli... The second edition, corrected},
  author={Winckelmann, Johann Joachim},
  year={1767},
  publisher={A. Miller\&T. Cadell}
}

@inproceedings{zhang2023adding,
  title={Adding conditional control to text-to-image diffusion models},
  author={Zhang, Lvmin and Rao, Anyi and Agrawala, Maneesh},
  booktitle={Proceedings of the IEEE/CVF international conference on computer vision},
  pages={3836--3847},
  year={2023}
}

@book{rodin2012rodin,
  title={Rodin on art and artists},
  author={Rodin, Auguste},
  year={2012},
  publisher={Courier Corporation}
}

@book{bush2012chinese,
  title={The Chinese literati on painting: Su Shih (1037-1101) to Tung Ch’i-ch’ang (1555-1636)},
  author={Bush, Susan},
  volume={1},
  year={2012},
  publisher={Hong Kong University Press}
}

@inproceedings{sivertsen2024machine,
  title={Machine learning processes as sources of ambiguity: Insights from ai art},
  author={Sivertsen, Christian and Salimbeni, Guido and L{\o}vlie, Anders Sundnes and Benford, Steven David and Zhu, Jichen},
  booktitle={Proceedings of the 2024 CHI Conference on Human Factors in Computing Systems},
  pages={1--14},
  year={2024}
}

@book{sullivan2008arts,
  title={The Arts of China, Revised and Expanded},
  author={Sullivan, Michael},
  year={2008},
  publisher={Univ of California Press}
}

\appendix
\section{Physical Requirements and Demonstration Feasibility for Ink Spiral}
\label{app:ink-physical-requirements}

\subsection{Information Sheet}

\textbf{Exhibit type.}  
Ink Spiral is a single-channel video installation intended for screen-based viewing. The work requires one large display and a standard power connection.

\vspace{0.5em}
\noindent
\textbf{Display.}  
The work requires one large screen for video presentation. A display in the range of 55--65 inches is recommended in order to ensure comfortable viewing in an exhibition setting. The display should support a minimum resolution of 1920 $\times$ 1080 and a refresh rate of at least 60\,Hz. Standard AC power is sufficient for operation.

\vspace{0.5em}
\noindent
\textbf{Spatial layout.}  
The installation may be presented within a compact footprint of approximately 1.5\,m $\times$ 1.5\,m to 2\,m $\times$ 2\,m. A viewing distance of approximately 1.5--2\,m is recommended. The work is suitable for individual viewing as well as small-group observation.

\vspace{0.5em}
\noindent
\textbf{Power requirements.}  
The installation requires only one standard power source for the display and playback device. A regular extension cord or power strip is sufficient. No specialized electrical infrastructure is needed.

\vspace{0.5em}
\noindent
\textbf{Lighting conditions.}  
A dim or controlled lighting environment is preferred in order to ensure clear visibility of the moving image. Strong direct light on the screen should be avoided.

\vspace{0.5em}
\noindent
\textbf{Sound environment.}  
If the video is presented with sound, a relatively quiet exhibition environment is preferred. If needed, the work may also be exhibited without external audio or adapted to headphone-based playback.

\vspace{0.5em}
\noindent
\textbf{Internet connection.}  
No Internet connection is required for presentation, as the work is a pre-rendered video installation.

\subsection{Feasibility of Demonstration at the Exhibition Site}

Ink Spiral is straightforward to install and technically compatible with standard exhibition conditions. The work does not require interactive equipment, network access, or complex setup. Its presentation only depends on one large display, one playback device, and access to standard AC power.

The installation is therefore feasible in most gallery, museum, or conference exhibition environments that can accommodate a small screen-based video work. Final presentation details may be adjusted according to venue conditions.
\end{document}